\documentclass[12pt]{iopart}
\usepackage{graphicx}% Include figure files
\usepackage{dcolumn}% Align table columns on decimal point
\usepackage{bm}% bold math
\usepackage{cite}%generates an in-text citation to the references associated with the keys in key_list
\begin{document}

\title{The dynamics of the HIV infection: a time-delay differential equation approach}

\author{Flora Souza Bacelar$^{a,b}$ and Roberto F. S. Andrade$^{a}$}
\address{
$^{a}$ Instituto de F\'{i}sica, Universidade Federal da Bahia,\\
40210-340 Salvador, Bahia, Brazil.\smallskip\\$^{b}$Instituto de Fisica Interdisciplinar y Sistemas Complejos (CSIC-UIB).\\ Campus Universitat Illes Balears, E-07122 Palma de Mallorca,
Spain.}
\ead{florabacelar@ifisc.uib-csic.es and randrade@ufba.br}

\author{Rita M. Zorzenon dos Santos}
\address{
Departamento de F\'{i}sica\\ Universidade Federal de Pernambuco\\
CEP 50670-901, Recife, Pernambuco, Brazil}
\ead{zorzenon@df.ufpe.br}
\date{\today}% It is always \today, today,

\begin{abstract}

In this work we introduce a differential equation model with time-delay
that describes the three-stage dynamics and the two time scales observed
in HIV infection. Assuming that the virus has high mutation and
rapid reproduction rates that stress the immune system throughout the
successive activation of new responses to new undetectable strains,  the
delay term describes the time interval necessary to mount new specific
immune responses. This single term increases the number of possible
solutions and changes the phase space dynamics if compared to the model
without time delay. We observe very slow transits near the unstable fixed
point, corresponding to a healthy state, and long time decay to the stable fixed
point that  corresponds to the infected state. In contrast to the results
obtained for models using regular ODE, which only allow for partial
descriptions of the course of the infection, our model  describes the
entire course of infection observed in infected patients: the primary
infection, the latency period and the onset of acquired immunodeficiency
syndrome (AIDS). The model also describes other scenarios, such as the very
fast progression to the disease and the less common outcome in which,
although the patient is exposed to HIV, he/she does not develop the
disease.

\end{abstract}

\pacs{02.30.Ks,02.30.Hq,87.18.Hf,87.19.Xx}
\vspace{2pc}
\noindent{\it Keywords}: Time delay, HIV, Virus strain
\maketitle

\newpage\

% \hrule height 2pt
%\tableofcontents
% \hrule height 2pt

\section{Introduction}

After almost three decades of research attempting to understand the dynamics
of HIV infection, many aspects of its underlying mechanisms remain
unrevealed. Understanding  the three-stages dynamics and  two
time scales has also challenged the scientific community over the past
decades. Despite the intrinsic differences between individuals, the
evolution of HIV infection follows a common pattern \cite{pantaleo} that
starts with the primary infection during the first weeks after
contamination and is characterized by an extensive dissemination of the
infection (large virus count), followed by a pronounced decline in the
virus count caused by the development of the virus-specific immune
response. Notwithstanding the decrease in the virus load, the system does
not recover completely from the infection after the primary infection and
very low concentrations of virus remain in the organism. This
second stage, called the latency period, is asymptomatic and varies from
patient to patient. The time-scale of the clinical latency period ranges
from months to several years, and  if untreated,  is
characterized by a progressive decrease in the number of HIV target cells,
the $TCD4^+$ cells. The third phase corresponds to the onset of acquired
immunodeficiency syndrome (AIDS) defined as the time when the T cell
counts reaches the order of $20-30\%$ of the concentration of the $TCD4^+$
cells in healthy individuals \cite{AIDS2005}.Without any treatment, the
patient dies from opportunistic diseases.

Among the efforts to understand the dynamics of  the disease, many
mathematical models have been proposed  to describe either specific
aspects of the HIV dynamics or the overall behavior of the evolution of
the infection. The attempts of the first two decades are reviewed in
details in \cite{nowak1, perelsonnelson} and references therein,  but
recent contributions still indicate an active  interest in the theme
\cite{ciupe2006, zhou2008, cuifang2009}. To date, most of the proposed
models have been  based on differential equations (ODE and
PDE) ( see for exemple \cite{nowak1,perelsonnelson,primaria1,hivterapia,landi2006,mukandaviere2006,barao2007}),
although discrete models have also been  considered \cite{MannionA,MannionB, rita, jafelice2009}. While most of the models proposed until 2001 were very successful in
describing specific aspects of the dynamics or the primary infection or
the latency period, none was able to describe the entire course of
the infection employing the same set of parameters. The first model that
described the  three-stage dynamics and two time scale as observed in
infected patients was a cellular automaton model put forward  to describe the
dissemination of infection in the target T cells located in the lymph
nodes \cite {rita}. Recently, Stilianakis and Schenzle
\cite{stilianakis2006} proposed an ODE approach to reproduce the entire
course of HIV infection.  The model describes the targeted infection of
$TCD4^+$ cells generating new variants of HIV that compete between
themselves with a capacity of increased reproducibility  within  the selected
variants that in turn leads to the immune system's loss of  control over
the infection. Using seven non-linear ODE' s and 22 parameters, the results
presented by the authors reproduce the entire course of the infection on
the time scale of days, corresponding to the very fast progression of the
disease. The authors claim that although the results presented do not
reproduce the typical course, they were able to reproduce (not revealed
in the paper) the whole spectrum of courses observed in infected
patients by varying the parameters. In a further discussion, they mention
that due to complications in the dynamics of the disease, the model loses
its applicability to the third phase of the infection.  According to the
results and discussions presented throughtout the paper the model seems to
reproduce the three stages  only for fast progressions of the disease (one
time scale) and fails to reproduce the two time scales (days and years)
observed in infected patients to describe the third phase.

Here we seek to describe the entire course of the HIV infection using a
unique set of parameters and a non-linear differential equation approach
that includes time-delayed terms. Our model follows the same assumptions as
the cellular automata model introduced by Zorzenon dos Santos and Coutinho
\cite{rita}:  HIV's high mutation and rapid replication rates and the
ability of the immune system to mount a new regular immune response to any
non-identified HIV resulting from the virus proliferation process. However,
since  a time interval ($\tau$) is necessary for the immune system
to mount any specific response developing from the necessary signaling among
the cells, the non-identified strain remaining during this time interval is
able to disseminate the infection throughout the target cells. In the
follow up, these newly infected cells would be able to produce new strains
(through mutations) that may not be recognized by the immune system (IS)
therefore, disseminating the infection during the same period of time.

The addition of this single delay term changes the dynamics of the model
in the phase space, generating trajectories that depend on its non-local
properties. These properties  change the transit of the trajectories near
the fixed points allowing for the description of the entire course of the
HIV with the two time scale and three-stage dynamics. Time -delay effects
are present in different (molecular, cellular or organ) mechanisms  of
various biological systems \cite{kuang} and recently,  other models with
time –delay terms  have been  introduced to describe HIV infection.
However, in these models, the time delay either represents a finite
incubation time before the release of new viruses by infected $TCD4^+$
cells \cite{hivdelay2criterio,hivdelaycelltocell} or describes the
delayed effects of the anti-retroviral drugs on the plasma viremia that
leads to the viremia decay \cite{cuifang2009, mukandaviere2006, hivdelay1, antiretroviraisanddelay}. Although these  ODE models have time
–delay terms, they do not reproduce the two time scales and three-stage
dynamics of the HIV infection as in the model we introduce.

%\end{Introduction}

\section{The model}

In order to facilitate the understanding of the model introduced herein,
let us briefly summarize the main assumptions of the CA model \cite{rita}.
The CA model describes the local interaction among target cells (T cells
and macrophages) and the virus in the lymph nodes, taking into account the HIV's
high mutation and rapid reproduction rates, as well as the fact that  the immune
system responds regularly to HIV as  to any other virus. The target
cells can be found in four states: healthy (\textbf{h}), infected
($\mathbf{A}$ and $\textbf{B}$) and dead (\textbf{d}) (or empty sites).
Healthy cells are the target cells that  may become infected. Infected-A
corresponds to the stage in which the infected cell disseminates the
infection during $\tau$ time steps (the period of time necessary to mount
the specific immune response) and infected-B corresponds to  the last
state of the infection in which the cells are recognized and  killed in
the next time step.  The dynamics of the interaction among automata is
defined by four rules: $(1)$ \textbf{h} cells become infected if  at
least one nearest neighbor is infected- \textbf{A}. Since  it is considered
that each new  infected- \textbf{A} that enters the system carries a new strain
of virus, $(2)$ asserts that infected -\textbf{A} cells spread the
infection into its vicinity during $\tau$ time steps, and afterwards the
infected-\textbf{A} cell  turns to a less infective-\textbf{B} stage.
Rule $(3)$ describes the turnover of \textbf{B} cells into \textbf{d} cells
(or empty sites) in the next time step. The regular  blood flow in the
system allows for the  natural replenishment of the target cells in the
lymph nodes, therefore  $(4)$ replaces \textbf{d} cells by \textbf{h} with
probability $p_{repl}$ or by infected-\textbf{A} cells with probability
$p_{repl}*p_{infec}$ , since we also consider that the empty sites may be
occupied by infected cells coming from other lymphatic compartments.
In the CA model, the fast dynamics of the primary infection corresponds to
a rapid increase in the number of infected cells through the cycle
\textbf{h}$\rightarrow$\textbf{A}$\rightarrow$\textbf{B}$\rightarrow$\textbf{d}$\rightarrow$\textbf{h}.
The slow dynamics of the latency period is related to the formation of the
spatial structures of infected cells. As shown in \cite{rita}, these
growing structures slowly compromise more and more cells, segregating and
trapping the healthy ones and leading to a reduction in $TCD4^+$ cell
counts. These structures are associated with syncytia (aggregation of
infected cells) formation, observed in HIV cultures and in the lymph nodes
of HIV patients.

Our model describes the time evolution of $TCD4^+$ cell density in
healthy, infected and dead states using a set of ordinary differential
equations. As in the CA model, we have differentiated the stages $A$ and $B$ in
the population of infected cells, but in turn there are also two differentiated
two stages in the population of healthy cells : $h_1$
corresponding to the healthy cells present in the tissue at the
beginning of the HIV infection, and $h_2$ corresponding to the  population of
new cells that enter the system when the infection has already set in. If
instead we consider only one type of healthy cells we will obtain the same
qualitative results, but shorter latency periods, as discussed bellow.  The
sum of all the variables (populations) remains constant over time and the
time delay terms would be included in the equations for infected-A and -B
cells to describe the time necessary for the IS to convert any new $A$
cell into a type $B$ cell.

Therefore, the model is described by the following set of differential equations:

\begin{eqnarray}\label{HIV1}
  \dot{h}_{1} &=& -k_{5}h_{1}\left( t\right) A^{p} -k_{6}h_{1}  B^{n} ,\nonumber \\
  \dot{h}_{2} &=& k_{3}d  -k_{5}h_{2}  A^{q}  -k_{6}h_{2}  B^{n}  ,\nonumber \\
  \dot{A} &=& -k_{1}A\left( t-\tau \right) +k_{4}d +k_{5}(h_{1}A^{p} + h_{2}
A^{q})+k_{6}(h_{1}+h_{2})B^{n},\nonumber \\
  \dot{B} &=& k_{1}A\left( t-\tau \right) -k_{2}B\left( t\right),\nonumber \\
  \dot{d} &=& -k_{3}d\left( t\right) -k_{4}d\left( t\right) +k_{2}B\left(
t\right).
\end{eqnarray}

where $p$ and $q$ represent the number of $A$ neighbors required to
convert healthy ($h_1$ or $h_2$) cells into newly infected-$A$ cells; $n$
is the number of $B$ neighbors necessary to infect  healthy cells ($h_1$
or $h_2$).  As in  \cite{rita}, we consider $p=1$, $n=4$ and $\tau=4$.
The $h_2$ population has no correspondent in the CA model but is
expected to be more resistant to HIV, requiring a larger number $q$ (when
compared to $p$) of $A$ neighbors to become infected. This assumption is
based on the fact that once the virus is detected and the specific immune
response is developed the system would respond faster to newly infected
cells, therefore more infected cells would be necessary to infected a
healthy one, once the infection has set in. The rate constants describe the
different transitions between states. Two of them can be directly related
to CA parameters: $k_{3}$ being the rate at which new  $A$-cells are
produced  is related to $p_{repl}$ mentioned above, and  $k_{4}$
corresponds to the rate at which newly infected cells enter the system(
$p_{repl}*p_{infec}$). The remaining rate constants, $k_{1}$, $k_{2}$,
$k_{5}$ and $k_{6}$ describe, respectively, the transitions from states
$A$ to the $B$, $B$ to $d$, and $h_1$ and $h_2$ to $A$.

As in the CA model \cite{rita}, values were chosen for some rate constants and parameters based on certain biological data  :  (i) the replenishment ability of the system $k_{3}~O(1)$ ;
(ii) a very low but finite probability  (1 in 10000 cells of the peripheral blood harbor the viral DNA) that some of the newly infected cells entering the system would come from other compartments $k_{4}<<1$; and (iii) since the responses to different virus can vary from a few days to 8 weeks,  as in the CA model, we adopted $\tau=4$.

The values of $k_i$ are not freely chosen, but obey  certain constraints that result from considering
: i) the coordinates of the fixed points ($FP=(\overline{h_1},\overline{h_2},\overline{A},\overline{B},\overline{d}$) representing densities that should remain in the $[0,1]$ interval; (ii) at the infected state the densities would respect the inequalities: $infected-A> infected-B>0$ and $d< infected-B$ ; and (iii)$p_{repl}>p_{inf}$ otherwise would replace the empty sites only by infected-A cells. Therefore, $k_{2}>k_{1}$, $k_5>k_1$, $k_{3}+k_{4}>k_{2}$, $k_{3}>k_{4}$.

%\end{The model}

\section{Results}

The system of equations (\ref{HIV1}) admits three FP solutions. Two,
$FP_{0}=(1,0,0,0,0)$ and $\widehat{FP_0}$ = $(0,1,0,0,0)$, represent the
healthy states and a third that represents the infected state $FP_{1} =
(0,\overline{h_2},\overline{A},\overline{B},\overline{d})$. To find its coordinates, it is necessary to solve
the following equation for $\overline{B}:$

\begin{equation}
R \overline{B}^{n}+ S \overline{B}^{n-1}+ T\left[\frac{k_2}{k_1}\right
]^{q-1} \overline{B}^q + U \left[\frac{k_2}{k_1}\right ]^{q-1}
\overline{B}^{q-1} + V = 0.\label{B1}
\end{equation}

Equation (\ref{B1}) is obtained by making the left-hand sides from the equations of system (\ref{HIV1}) equal zero, as well as considering the condition of normalization. Once equation (\ref{B1}) has been obtained, it is then possible to describe the behavior of the infected B-cells, in which coefficients R, S, T, U and V are functions of the parameters of the model. Depending on  the value of $q$ ( integer or not), (\ref{B1})
becomes either a polynomial or a transcendental equation. However,
for integer $q\in [0,4]$, the resulting analytical expression for
$\overline{B}$ is very complex and has no practical
application to the analysis of $FP_{1}$. Therefore,
(\ref{B1})is always solved by numerical methods.

$FP_{0}$ and $\widehat{FP_{0}}$ are both unstable. $FP_{0}$ is reached
only for initial condition with  $h_1$=1, while few heteroclinic orbits
converge to $\widehat{FP_{0}}$ along the direction of its attractive
manifold. To analyze the stability of the $FP_1$, we follow the standard
procedure using the equation for the eigenvalues of the Jacobean matrix
$J$, $M(\lambda)$=$det(J-\lambda I)$=0. For  $\tau>$0, $M(\lambda)$=0
becomes a transcendental equation since it contains terms that depend on $
e^{-\lambda\tau}$. Irrespective of the value of $\tau$, we always obtain
an identically vanishing eigenvalue ($\lambda_0$), related to the
conservation of the number of cells.

When $\tau$=0 there is no delay in mounting the specific immune response and its spectral equation is polynomial. For example,  if we choose $k_{1}$=0.16, $k_{2}$=0.2, $k_{3}$=0.68,
$k_{4}$=$2.5\times10^{-5}$, $k_{5}$=0.6, $k_{6}$=0.1, $n$=4, $p$=$q$=1, we
obtain $FP_{1}$=(0, 0.266,0.361,0.289,0.085), while the spectrum is given
by $(\lambda_0,\lambda_1,\lambda_{2,3},\lambda_4)$=(0,-0.705,-0.196$\pm$0.215i,-0.217).
Since $\overline{h_1}=0$, $\lambda_4$ describes only the decay of $h_1$ to
0; $\lambda_{1}$ indicates a fast decay to a two-dimensional space in which
the slow decay towards $FP_1$ , actually described by $\lambda_{2,3}$,
takes place. In other words, as observed in the CA model \cite{Guillermo}
the absence of time delay in mounting the specific response would favor
the spread of the infection and shorten the latency period. If instead of
unity we adopt $q=1.13$, the decaying dynamics slow down as
$Re(\lambda_{2,3})=-0.169$, i. e.,  the effect of shortening the
latency period  can be reduced if we increase the resistance of the new
incoming healthy cells.

\begin{figure}[tbp]
\centering
\includegraphics*[width=10cm,angle=0]{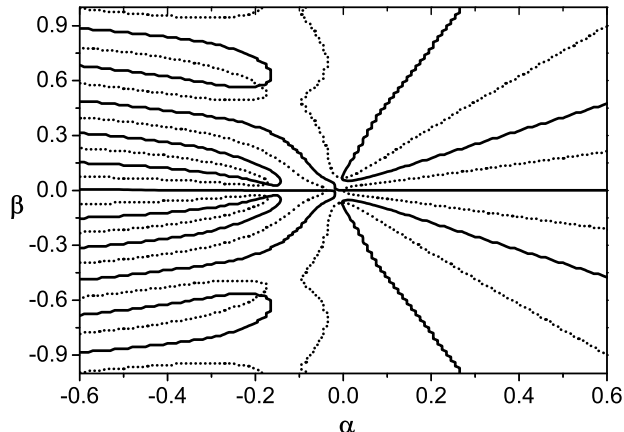}\vspace{-1.0cm}
\caption{Location of some eigenvalues in the $(\alpha,\beta)$ plane,
corresponding to the intersections of the nullclines $u(\alpha,\beta)=0$
(solid) and $v(\alpha,\beta)=0$ (dotted). The main features are:
$u\equiv0$ when $\beta=0$; there are two real eigenvalues for $\alpha=0$
$(\lambda_0)$, and $\alpha<0$ $(\lambda_4)$;there is  a pair of complex eigenvalues
($\lambda_{2,3}$) close to the origin; and an absence of eigenvalues with
$\alpha>0$. }
\end{figure}

For $\tau>0$, the complex eigenvalues are obtained employing  the Newton-Raphson (NR) procedure, using the notation  $\lambda =\alpha + i \beta$ where $\alpha=Re(\lambda)$
and $\beta=Im(\lambda)$,  $M(\lambda)= u(\alpha,\beta) + i
v(\alpha,\beta)$.   The eigenvalues correspond to the points where the
nullclines of $u(\alpha,\beta)=0$ and $v(\alpha,\beta)=0$ intersect in the
$(\alpha,\beta)$ plane, as illustrated in Figure 1. Graphs are very
helpful for finding roots of $M(\lambda)=0$, since  they indicate the
initial values for the NR procedure. The overall picture remains
essentially the same if we enlarge the region of the $(\alpha,\beta)$
plane in which the nullclines are drawn. Note that the nullclines cross
each other only in the negative $\alpha$ region and  $FP_{1}$ remains
stable. The several nullcline branches, when $\alpha<0$, indicates that
both $u$ and $v$ oscillate and that there are an infinite number of
solutions to $M(\lambda)=0$. The influence of $\tau$ on the decay rate
influences only the eigenvalues with the smallest real part, i.e.,
$\lambda_{2,3}$. The stability analysis shows that, when $\tau$ is
increased from 0 to 4, the decay time to $FP_1$ is amplified by factor
$7$  for $q=1.13$, and factor $ 5.4$ for $q=1.0$. All results of
the stability analysis for $\tau=0$   are checked by measuring the slope of
the amplitude of  the decaying solution to $FP_1$ that coincides with
$Re(\lambda_{2,3})$.

\begin{figure}[tbp]
\centering
\includegraphics*[width=8cm,angle=-90]{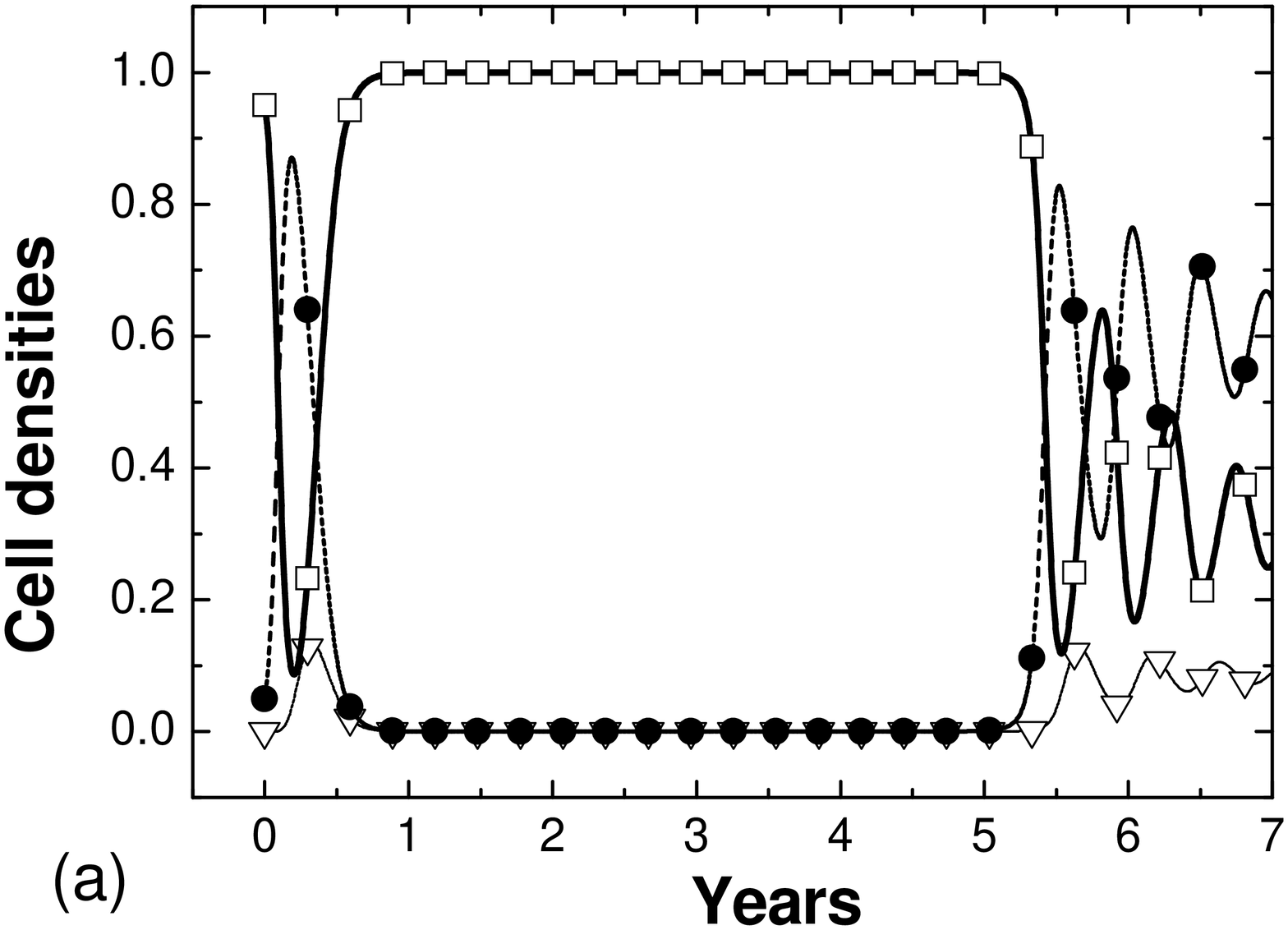}\\
\includegraphics*[width=8cm,angle=-90]{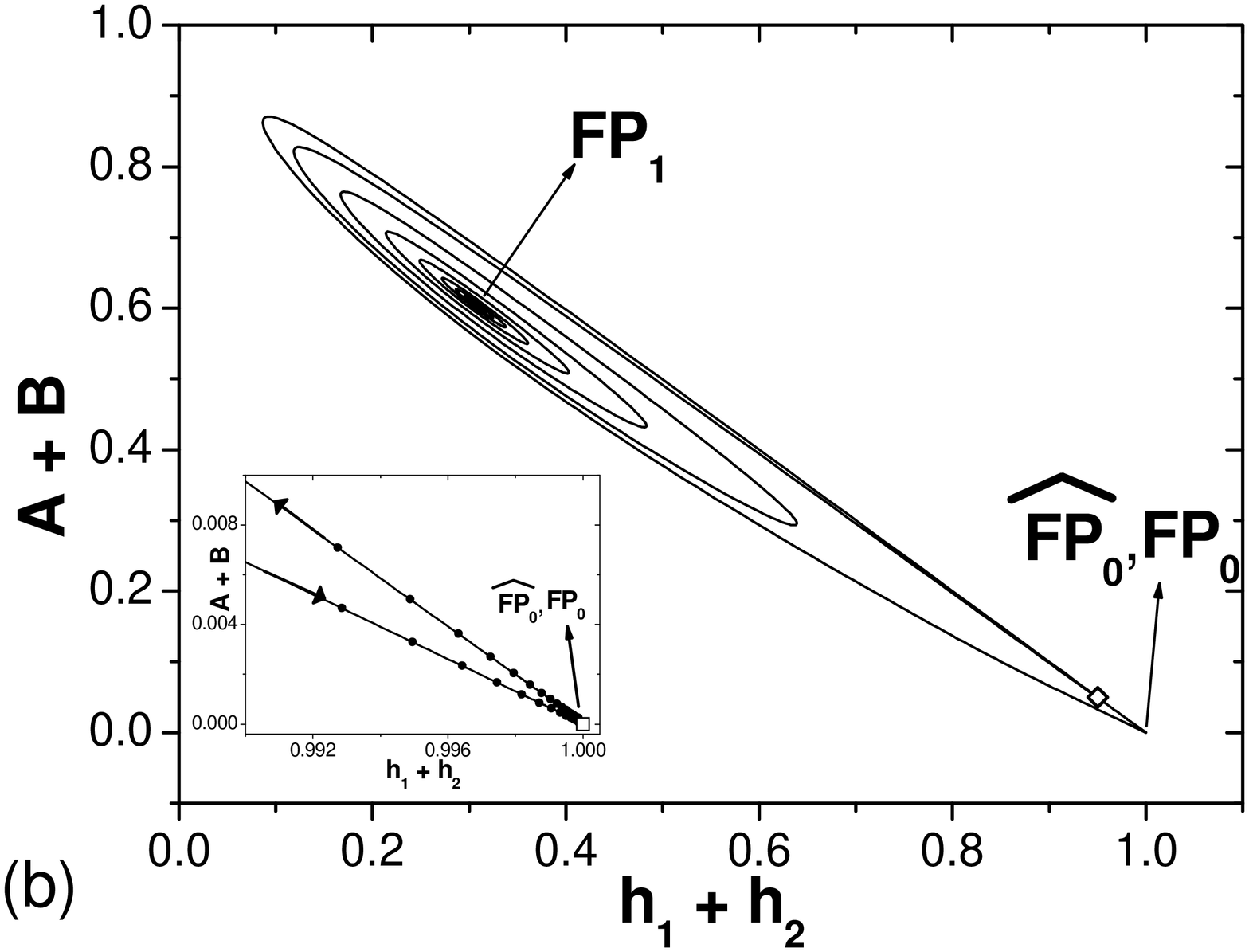}\\
\caption{(a) Time evolution of HIV infection showing  three distinct phases.
Squares, circles and triangles represent $h_1+h_2$, $A+B$ and $d$ cells,
respectively. The parameter values and initial conditions adopted are: $k_{1}$=0.163, $k_{2}$=0.228, $k_{3}$=0.65, $k_{4}$=3.25$\times$10$^{-5}$, $k_{5}$=0.606, $k_{6}$=0.02, $n$=4, $p$=1, $q$=1.13, $h_1$=0.95, $A$=0.05. (b) The same
orbit in a plane projection of phase space. Diamond indicates the initial
point. Inset: details of the first transit close to $\widehat{FP_0}$. The
trajectory first lies in the attracting manifold, but later it is pushed
away from $\widehat{FP_0}$ along the repulsive manifold. Heteroclinic
orbits occur if $\widehat{FP_0}$ is reached.}
\end{figure}

If we vary the parameters respecting the conditions discussed at the
beginning of this section, we note that the coordinates of FP1 change in a
continuous manner and  $FP_{1}$ remains a stable fixed point. Furthermore,
by fixing the values of the rate constants, we find that the attracting properties of $FP_{1}$
 weaken as  $q$ and $\tau$ increase. These two parameters of the
model contribute to reducing the velocity at which the trajectories
decay. In other words according to this model the duration of the latency
period is determined by the slow dynamics on the trajectories close to the
stable fixed point, which in turn is controlled by parameters $q$ and
$\tau$. By increasing the value of $q$  we require a larger number of
infected-$A$ cells in order  to infect healthy cells $h_2$, indicating
that an increase in the resistance of the incoming target cells also
increases the latency period. On the other hand, greater the $\tau$, the
greater the latency period or the individual' s  survival lifetime.
Nevertheless, these features are not sufficient in themselves  to account by themselves
for the observed increase in the time-scale associated with the latency
phase. It also depends on the presence of heteroclinic orbits
generated by  the time delay term,  uncovered by numerically integrating
the set of equations (\ref{HIV1}) using a fourth order Runge-Kutta method
adapted to the time-delay approach.

As in the case of the FP analysis, a detailed investigation of all
possible trajectories of the system (\ref{HIV1}) for large regions of
parameter space has been carried out. For the sake of clarity, here we
report only the results  that  illustrate the  typical patterns
obtained in different regions of the parameter space. Since $FP_{0}$ is unstable, almost all trajectories starting in its neighborhood are pushed away. When  $\tau=0$ the system evolves
rapidly to the infected state ($FP_1$) and the onset of AIDS occurs a few
weeks after the infection. This  would correspond to cases where the disease progresses rapidly
, which  in general  is observed in HIV
patients who have been contaminated under immune-depression conditions,
caused by malnutrition or other diseases that compromise $TCD4^+$ cells
(e.g. tuberculosis). If we  increase the value of $\tau$ from 0 to 4, the
trajectory is deviated from $FP_1$ towards the neighborhood of
$\widehat{FP_{0}}$, where the slow transit leads to the a steady increase
of latency period.

Figure 2(a) summarizes the main  results obtained for this model: the
appropriate description of the three-stage dynamics observed in infected
individuals who have not been submitted to drug therapies. In contrast to
other ODE models, here, using a unique set of parameters, we were able to
reproduce the primary infection and  a large range of  latency periods
(plateau in Figure 2a) that vary from weeks to years. Moreover, a
plateau is obtained on the T cell counts when  the trajectory on the phase
space goes towards the region of very slow dynamics, close to
$\widehat{FP_{0}}$ (see Figure 2b and inset). As previously mentioned, the
deviation of the trajectory in the phase space to the slow transit region
largely depends on the values of $q$ and $\tau$. Since an increase of
these parameters leads to a decrease in $Re(\lambda_{2,3})$ , the return
of the phase space trajectory  towards $FP_{1}$ proceeds at a much slower
pace, thus increasing the length of the latency period.

With regard to the estimates of the latency period, we observe that at the
end of the plateau, the healthy $TCD4^+$ cell counts oscillates for some
time until it stabilizes below the threshold associated to the onset of
AIDS. Since the latency period is defined as the time between
the end of the primary infection and the onset of AIDS,  we may  make a
second estimate of the latency period that will be larger than that
roughly estimated from the plateau. The parameter values used in Figure 2
were chosen so as to maximize the length of the latency phase. However,
the same qualitative pattern, with a well-defined latency period, is
observed  for a finite region of the parameter space. Usually, the
three-phase pattern remains stable to changes $\sim1-2\%$ in parameter
values of  the reported sets. Besides this, if no distinction is made between
$h_1$ and $h_2$, ($p=q$), similar patterns are obtained, albeit with a
smaller plateau. For other regions of the space parameter, the decay of the
trajectory to $FP_1$ may proceed with less defined slow transits close to
$\widehat{FP_{0}}$ and therefore lose the clearly defined
latency period.

Different patterns other than those of the three-stage dynamics may be obtained
for other parameter values not considered here. Without going into
detailed description, a brief comment follows on the most
important patterns obtained by changing the values of $q$ and $\tau$.  For low
values of $q$, the plateau is reduced, however, if it is increased beyond
1.18, we observe that all trajectories starting in a neighborhood of
$FP_0$ converge to $\widehat{FP_{0}}$. These heteroclinic trajectories,
linking two unstable FPs that are not related to the infected state,
describe the situation well documented in the
literature, where the patient has contact with the virus but does not
develop the disease. Moreover, for $q=1.13$ and $\tau=3$, the latency
phase is still short, while, for $\tau=5$, a sustained oscillation pattern
is found due to the existence of stable limit cycles around a locally
stable $FP_{1}$. In such a case, two attracting stable sets are present. In
other words for $\tau>4.5$ the monotonic behavior disappears.

Therefore, it is possible 'grosso modo'  , to group the different
trajectories into the following groups: i) rapid decay to $FP_{1}$; ii)
decay to $FP_{1}$ through a latency phase, whose duration will depend on
the values of $q$ and $\tau$; iii) trajectories converging to
$\widehat{FP_{0}}$ for $q$  sufficiently large ($>$1.18); iv) stable
limit cycle around $FP_{1}$ with a large amplitude, when $\tau$ is larger
than 4.5

Figure 3 illustrates the average values of the cell counts resulting from
the integration of  (\ref{HIV1}) for $N_t=200$ trajectories. The initial
conditions were randomly chosen from the very  close neighborhood of $FP_0$
for the same parameter values used in Figure 2. This assumption may be
interpreted as if each of the 200 individuals had been exposed to
different initial viral loads. Similar results are obtained if slight
fluctuations of the parameters are permitted to characterize different
individuals of the simulated population. When we compare the results
obtained  to its corresponding counterparts obtained with the CA model
\cite{rita}, we observe the same overall  dispersion pattern  in both.
However, in the present case the average dispersion is lower in the
primary infection and after the onset of AIDS, but as obtained with the CA
model it increases greatly during the latency phase. The large error bars obtained during the latency period indicate a large range of variability on the development of the disease
among the 200 individuals (trajectories).

\begin{figure}[tbp]
\begin{center}
\includegraphics*[width=8cm,angle=-90]{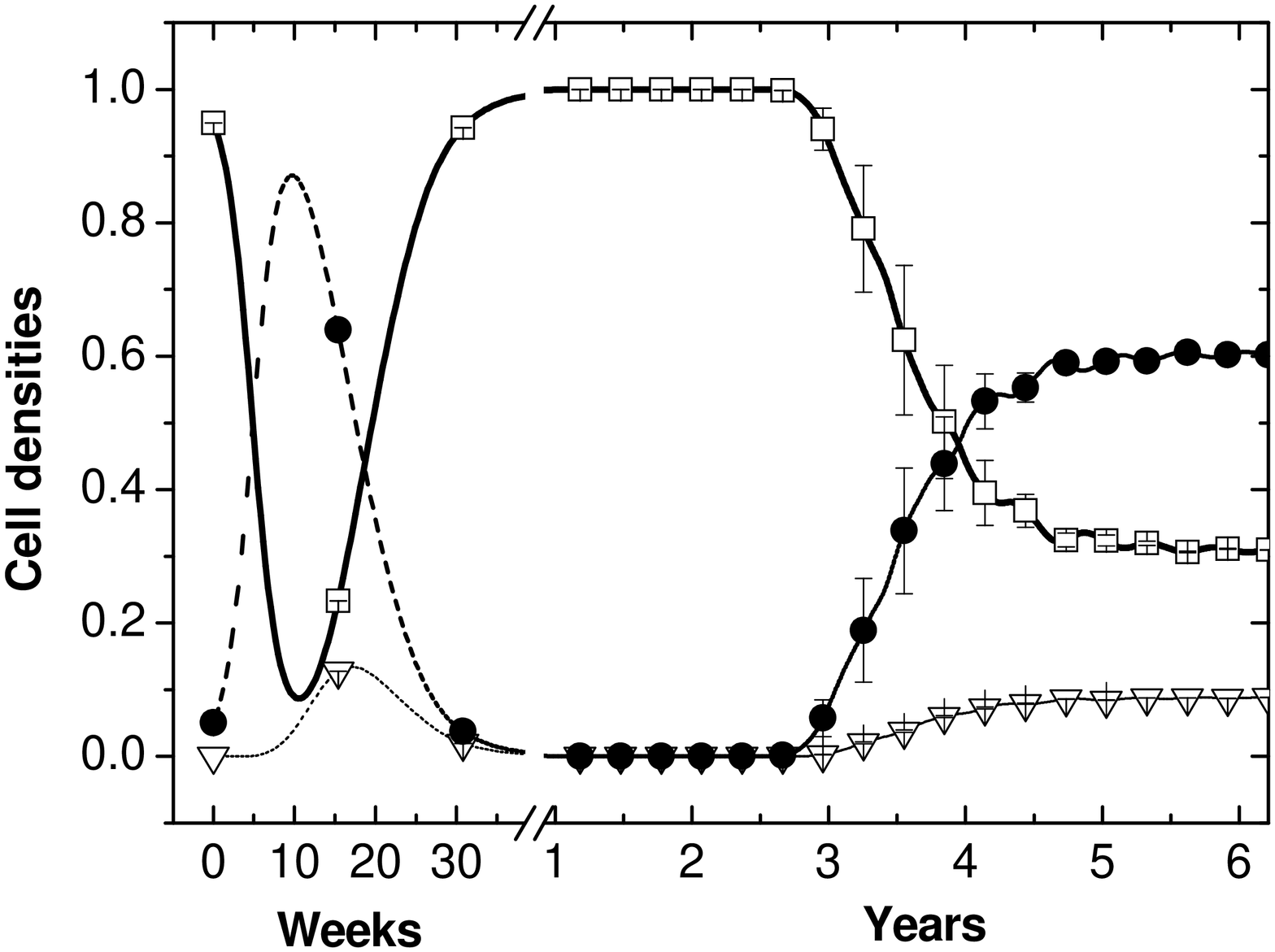}
\end{center}
\caption{Average values of healthy, infected and dead cell counts resulting from 200 trajectories using different  initial conditions starting in a neighborhood of $FP_0$.
The parameter values are the same as those used in Figure 2, as are  the symbols used to describe different concentrations . The error bars
indicate the standard deviation.}
\end{figure}

The results of this study show that a set of ordinary differential
equations with time delay terms describing the retard on mounting the
specific response, is able to provide a description of the entire course of
the HIV infection using a single parameter set. To our knowledge, this has
not been achieved previously by any other similar approach. As the
model is essentially based on the rules of a CA model, our results
indicate that these rules capture the main elements and mechanisms
underlying the dynamics of the infection of $TCD4^+$ cells. In the CA model,
spatial localization is very important for the emergence of distinct
time-scales. In the model presented here, with no spatial dependence, the
individual phases are found to be associated with particular structures in
the phase space, which are induced by the time-delay terms. These terms
describe the interplay amongst new generations of infected cells (strains)
and new specific responses and increase the number of possible classes of
solution. Although the number of different solutions is not as broad as
those of systems with explicit spatial dependence, it is sufficient to
generate phase space structures that drive the system to regions of very
slow dynamics.  A more detailed discussion of all types of solution for
system (\ref{HIV1}) will be published elsewhere.

Acknowledgements: This work was partially supported by the following
Brazilian funding agencies: CNPq, CAPES and FACEPE (grant PRONEX/FACEPE
EDT 0012-05.03/04 and PRONEX/FACEPE APQ 0203-1.05/08.). RMZS would also like to thank the hospitality of KITP -UCSB during the conclusion of this research paper, which was therefore  supported by the National Science Foundation under grant No. NSF PHY05-51164.

\end{document}